\documentclass[checkin,showpacs,aps,prx]{revtex4-1}

\newcommand{\bn}{\begin{enumerate}}
\newcommand{\en}{\end{enumerate}}
\newcommand{\ba}{\begin{eqnarray}}
\newcommand{\ea}{\end{eqnarray}}

\usepackage{graphicx,color}

\newcommand{\be}{\begin{equation}}
\newcommand{\ee}{\end{equation}}

\newcommand{\et}{{\it et al. }}
\newcommand{\ete}{{\it et al.}}

\def\prl{{ Phys. Rev. Lett. }}

\def\prb{{ Phys. Rev. B }}

\begin{document}

\newcommand{\clr}{\color{black}}








\title{ First-principles and model
simulation of all-optical spin reversal}


\author{G. P. Zhang,  Z. Babyak and Y. Xue}

 \affiliation{Department of Physics, Indiana State University,
   Terre Haute, IN 47809, USA }

\author{Y. H. Bai}

\affiliation{Office of Information Technology, Indiana State
  University, Terre Haute, IN 47809, USA }

\author{Thomas F. George}

\affiliation{Office of the Chancellor and Center for Nanoscience
  \\Departments of Chemistry \& Biochemistry and Physics \& Astronomy
  \\University of Missouri-St. Louis, St.  Louis, MO 63121, USA }

\date{\today}

\begin{abstract}
{All-optical spin switching is a potential trailblazer for information
  storage and communication at an unprecedented fast rate and free of
  magnetic fields. However, the current wisdom is largely based on
  semiempirical models of effective magnetic fields and heat pulses,
  so it is difficult to provide high-speed design protocols for
  actual devices. Here, we carry out a massively parallel
  first-principles and model calculation for thirteen spin systems and
  magnetic layers, free of any effective field, to establish a simpler
  and alternative paradigm of laser-induced ultrafast spin reversal
  and to point out a path to a full-integrated photospintronic device. It
  is the interplay of the optical selection rule and sublattice spin
  orderings that underlines seemingly irreconcilable
  helicity-dependent/independent switchings. Using realistic
  experimental parameters, we predict that strong ferrimagnets, in
  particular, Laves phase C15 rare-earth alloys, meet the telecommunication
  energy requirement of 10 fJ, thus allowing a cost-effective
  subpicosecond laser to switch spin in the GHz region.  }
\end{abstract}




 \maketitle

\section{INTRODUCTION}



Advanced material engineering and ultrafast laser technology
revolutionize the way that information is stored and transmitted
\cite{bader}.  Over a half century, switching magnetic moments
exclusively relies on magnetic fields, but now multiferroics allows
the electric field to control spins \cite{mundy}.  Spins can also be
manipulated by correlated spin–-charge quantum excitations \cite{li}
and intense terahertz transients
\cite{kampfrath11,kampfrath13,kubacka,baierl}. Spin-orbit coupling
adds a new dimension to control spin currents \cite{lesne}. Very
recently, 55-fs spin canting in Fe nanoparticles was discovered
\cite{ren}.  Remarkably a single laser pulse is capable of switching a
quantum spin from one orientation to another \cite{stanciu}, free of a
magnetic field. This all-optical spin switching (or AOS) immediately
ignited the entire community of ultrafast magnetic storage and
information communication
\cite{sabineprb,sabineapl,sabineprb14,hass,hassdenteufel2015}, but
results are much more complex.  The switching in ferrimagnetic GdFeCo
was found to be helicity-dependent \cite{stanciu}, but when the laser
fluence was above a particular threshold, helicity-dependent spin
switching (HDS) transitions to helicity-independent switching
(HIDS) \cite{ostler}. However, such transition is not seen in other
ferrimagnets \cite{mangin,hass,sabineprb14,sabineapl} nor in
ferromagnets \cite{lambert}.  A stronger laser does not lead to HIDS
but only demagnetizes the sample.  These paradoxically contradictory
results challenge our understanding and are difficult to reconcile.
Over the years, the explanation progresses from the inverse Faraday effect
\cite{stanciu,vahaplarprl}, Raman scattering \cite{gr,pop}, magnetic
circular dichroism \cite{khorsand}, pure heating \cite{ostler},
and sublattice spin exchange \cite{mentink}, to ultrafast exchange
scattering \cite{bar}, with new theories emerging in ferromagnets
\cite{gorchon,gr2,corn}. This raises a serious question whether a big
picture is missing from the existing theories
\cite{mplb16}. Furthermore, no theory ever addresses a design protocol
for future photospintronic devices based on AOS technology \cite{miller}.

In this paper, we establish an alternative and simpler paradigm for
laser-induced all-optical spin reversal and establish a path to
future applications. We carry out an extensive time-dependent
first-principles and model calculation for thirteen carefully selected
spin and layer systems. Different from prior studies, our theory
does not invoke an effective magnetic field or a heat pulse, thus
reflecting the experimental situation better. We show that the
helicity-dependent AOS is the manifestation of the optical selection rule,
a finding that is corroborated by the first-principles
results. Sublattice spins provide additional degrees of freedom to
control spin reversal. However, for a weak ferrimagnet, the selection
rule is still operative, so the switching is helicity-dependent. A
sudden change occurs in a strong ferrimagnet, where the sublattice
spins differ a little in their magnitude and switching becomes
helicity-independent. We construct a phase diagram for the entire spin
reversal. Using the experimental parameters \cite{chen}, we show that
in the strong ferrimagnet limit, the energy consumption is already
below the technological requirements. We find that Laves phase C15
rare-earth alloys are ideal candidates for future spin switching in
the GHz region.

\section{Theoretical formalism}

There are several attractive theories available, but most of them
introduce an effective magnetic field or a heat pulse, whereas
experimentally no magnetic field is applied. We see that there is room
for improvement.  We employ two complementary theories: one is the
first-principles method, and the other is a model simulation.  Such a
joint study is necessary, as seen below, in that it allows us to flexibly
investigate different aspects of all-optical spin reversal and cross
check the results, so we can develop a simple and more complete
picture for AOS. Different from prior theories, none of our theories
needs either an effective magnetic field or a heat pulse.  So our
theories are closer to the experimental reality, and present an
alternative to existing theories which are based on the
Landau-Lifshitz-Gilbert Bloch formalism. We will draw connections
with those prior theories, whenever possible.

\subsection{Time-dependent first-principles calculation}

\newcommand{\ik}{nk}

In our first-principles studies, we first solve the Kohn-Sham equation
(in atomic units) self-consistently, \be
[-\nabla^2+V_{Ne}+V_{ee}+V_{xc}^\sigma]\psi_{\ik}(r)=E_{\ik}
\psi_{\ik} (r). \label{ks} \ee where the terms on the left side
are kinetic energy, electron-nuclear attraction energy, Coulomb and
exchange correlation, respectively.  $\psi_{\ik}(r)$ and $E_{\ik}$ are
the eigenstates and eigenenergies at the $k$ point for band $n$.  We use
the full-potential augmented plane-wave method as implemented in the
Wien2k code \cite{wien}, where the spin-orbit coupling is also
included.  The dynamic simulation starts with the Liouville equation,
\be i\hbar \frac{\partial \rho}{\partial t}=[H_0+H_I,\rho], \ee where
$\rho$ is the density matrix, and $H_0$ is the unperturbed system
Hamiltonian. $H_I$ is the interaction between the system and laser
field: $H_I=\sum_{k;i,j} {\bf P}_{k; i,j} \cdot {\bf A}(t)$, where
${\bf P}_{k;i,j}$ is the momentum matrix element between states $i$
and $j$ at $k$, and ${\bf A}$ is the vector potential with
amplitude $A_0$.  For all the first-principles calculations below, we
use the vector field potential amplitude $A_0$ in the unit of $\rm V
fs/\AA$.  Once we solve the Liouville equation, we compute the spin
expectation value by the trace ${\rm Tr}(\rho S_z)$.

\subsection{Model simulation}

In our model simulation, we adopt a thin slab, with two monolayers
along the $z$ axis and 21 lattice sites along the $x$ and $y$ axes.
Spins are arranged orderly in a simple cubic structure, thus removing
any ambiguity in spin configuration.  We verify that our system is
large enough that the finite size effect is small. When we construct
our model, we are mindful that it can not include every detail in a
sample; otherwise, the problem would become intractable.  With this in
mind, we construct our model Hamiltonian as \cite{epl15,epl16,mplb16}
\be H=\sum_i \left [\frac{{\bf p}_i^2}{2m}+V({\bf r}_i) +\lambda {\bf
    L}_i\cdot {\bf S}_i -e {\bf E}(t) \cdot {\bf r}_i\right
]-\sum_{ij}J_{ex}{\bf S}_i\cdot {\bf S}_{j}, \label{ham} \ee where the
terms on the right side are the kinetic energy, potential energy,
spin-orbit coupling, interaction between the laser field and the
system, and the Heisenberg exchange interaction between the
nearest-neighbor sites.  A similar form is often used for magnetic
multilayers \cite{hs,manchon2009}.
${\bf L}_i$ and ${\bf S}_i$ are orbital and spin angular
momenta at site $i$, respectively, and $J_{ex}$ is the exchange
integral in units of eV/$\hbar^2$.  Since each site contains one spin,
as a standard practice, we use the same index $i$ to denote both the
spin and atomic site.  The nearest-neighbor spins are coupled either
antiferromagnetically or ferromagnetically. Ferrimagnets have two
sublattices, $S_z^a$ and $S_z^b$.

Our model contains four minimum conditions for
     spin reversal: \bn
\item[(i)] A channel for the laser to transfer the energy and angular
  momentum into the system. \item[(ii)] A transient increase of the
  orbital angular momentum \cite{john}. \item[(iii)] Emergent
  spin-orbit torque \cite{epl16}. \item[(iv)] Spin-spin
  interaction \cite{ostler}.  \en

One can show easily that with any one of them missing, spin switching
founders. To make connections with prior theories, we should point out
that the inverse Faraday effect \cite{stanciu} is intrinsically
connected to the spin-orbit coupling in Eq. (\ref{ham}), and both
Raman scattering \cite{gr,pop} and magnetic circularly dichroism
\cite{khorsand} are included through the first four terms in the
equation, while the sublattice exchange interaction \cite{mentink} and
scattering \cite{bar} are included in the last term of the equation.
Ostler \et \cite{ostler} essentially replaced the first three terms by
a phenomenological heat pulse.  The major difference between our work
and reference \cite{pop} is that they worked with the
wavefunction, so the spin-orbit torque was hidden behind the
convoluted wavefunction. In our theory, we work with operators
directly, so it is easier to reveal the role of the spin-orbit torque
in spin reversal.  Our theory \cite{epl15} also recovers the results
by Pershan {\it et al.}  \cite{pershan1966}.  Spin-orbit torque is
also similar to the spin orbit-induced torque by Manchon and Zhang
\cite{manchon2009}.  The only difference is that their driving field
was current and in our case, we have a laser field.  Our Hamiltonian
is a quantum mechanical many-body Hamiltonian.  Such a model is
difficult to solve exactly, and approximations have to be made.

We solve Heisenberg's equation of motion numerically for each operator
of interest under the influence of a laser field within the
Hartree-Fock approximation. The validity of this approximation is
checked by comparing our results with the experimental ones.  The
exchange interaction is $J_{ex}=0.1{\rm eV}/\hbar^2$, and the laser
pulse duration is $\tau=240$ fs.

\section{Results and discussions}

\subsection{Optical selection rule}
\begin{figure}
\includegraphics[angle=0,width=0.8\columnwidth]{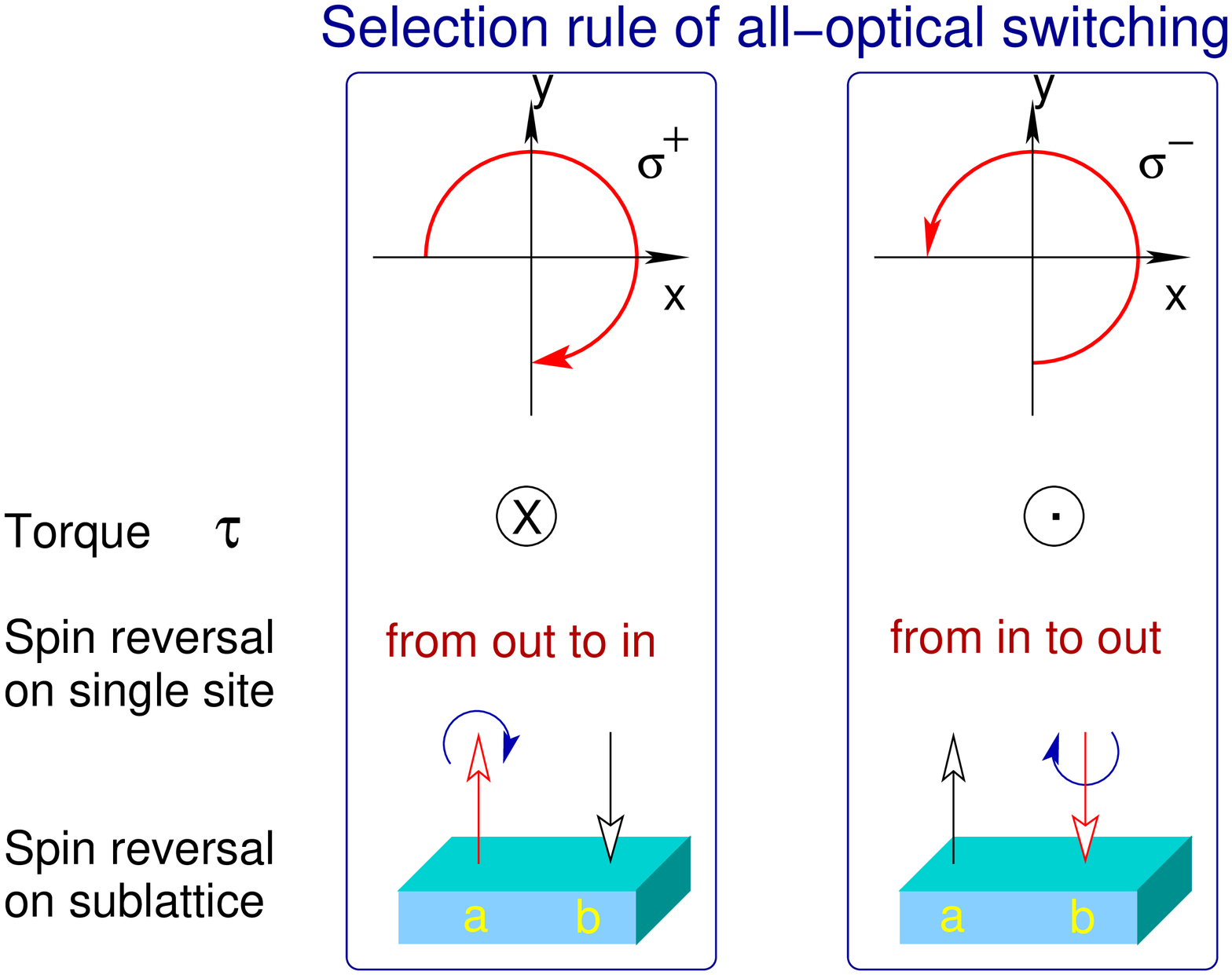}
\caption{Selection rule for all-optical spin switching. The light
  helicity determines the direction of the spin-orbit torque
  $\tau$. (Left) For a single spin, right-circularly polarized light
  ($\sigma^+$) rotates the spin from out of the page into the page.
  (Right) $\sigma^-$ light does the opposite. (Bottom) For a system
  with two sublattices $a$ and $b$, $\sigma^-$ and $\sigma^+$ switch
  different sets of spins.  $\sigma^+$ switches a spin from up to
  down, while $\sigma^-$ switches a spin from down to up.  }
\label{fig0}
\end{figure}

 To start with, we note that all-optical spin reversal is an optical
 process and must follow the dipole selection rule.  Consider a laser
 field propagating along the $-z$ axis toward a sample surface
 \cite{epl15,mplb16} (see Fig. \ref{fig0}), \be {\bf E}(t)=E_0 {\rm
   e}^{-t^2/\tau^2} (\pm \sin(\omega t) \hat{x}+\cos(\omega t)
 \hat{y}), \label{efield}\ee where $E_0$ is the laser field amplitude in the unit of
 $\rm V/\AA$ (not to be confused with the vector potential $A_0$
 above), $\omega$ is the carrier frequency, $\tau$ is the laser pulse
 duration, and $\hat{x}$ and $\hat{y}$ are the unit vectors along the
 $x$ and $y$ directions, respectively. $+(-)$ refers to right-
 (left-) circularly polarized light, $\sigma^{+(-)}$.  In atoms, any
 spin states are characterized by the total angular momentum quantum
 number $J$ in the presence of spin-orbit coupling; in solids, the
 rule is still there but manifests itself through the optical
 transition matrix elements at every crystal momentum point in the
 reciprocal space.  For right- and left-circularly polarized light
 $\sigma^+$ and $\sigma^-$, $J$ changes as  \cite{prb08}
\begin{equation}
\Delta J= \left \{%
 \begin{array}{lll}
  +1~~ &\hspace{1cm }(\sigma^+) & \hspace{1cm} \uparrow \Longrightarrow \downarrow\\
  -1~~&\hspace{1cm} (\sigma^-)  & \hspace{1cm} \downarrow \Longrightarrow \uparrow
 \end{array}%
\right.
,
\end{equation}
where the double line arrows emphasize the angular momentum passage
between the spin and orbital degrees of freedom.

To visualize how the spin reversal happens, Fig. \ref{fig0} illustrates the helicity
dependence of spin reversal in the $x-y$ plane. On the left side of
the figure, we have right-circularly light, where the electric field
rotates clockwise and its induced spin-orbit torque $\tau_{soc}$
\cite{epl16} follows the normal right-hand rule. If we curl our
fingers along the light helicity direction, the thumb points in the
direction of the torque. In this case, it points into the page. If the
original spin points out of the page, under the influence of this
torque, it will be reversed into the page. But if the spin already
points into the page, then there is no effect on this spin. If we
choose left-circularly polarized light (see the right side of
Fig. \ref{fig0}), the situation is reversed and $\tau_{soc}$ points
out of page.  This rule is very powerful and allows us to figure out
how the spin reverses. The bottom panel shows that the thin film has
two spin sublattices, $a$ and $b$. Suppose the spin on $a$ points out
of the plane of the film and that on $b$ into the plane. If the
$\sigma^+$ laser comes down on the film, only the sublattice spin $a$
(in red) is affected. The effect on sublattice spin $b$ is through the
exchange interaction. If we use $\sigma^-$, then the spin on
sublattice $b$ is affected.  However, the selection rule only provides
a possibility to switch spins, but can not give a definitive answer
whether the reversal actually occurs. This requires a first-principles
calculation.

\subsection{Time-dependent Liouville density-functional study of helicity dependence}

Chimata \et \cite{chimata} employed the first-principles method, but
their switching in Gd-Fe alloys was simulated via a model
\cite{ostler}. Another approach also appeared \cite{berritta}, where
the calculation was static.  Time-dependent density functional theory
has been employed to investigate an ultrafast demagnetization field
\cite{krieger,elliot2016,simoni,stamenova} but not for spin reversal.
We carry out an extensive density functional calculation and
time-dependent Liouville simulation \cite{np09} under circularly
polarized light.  This method slightly differs from the traditional
time-dependent density functional theory, where our time propagation
is done through the Liouville equation and electron excitation is
described by the density matrix.  We employ three element ferromagnets
(Ni, Fe and Gd) and one alloy (CoPt), with different structures (fcc,
fct, hcp and monolayer) and four sets of laser parameters with left-
and right-circularly polarizations, together with eight Laves phase
(C15) rare-earth intermetallic compounds (see below). The details of
the structural and magnetic information are given in the Supplementary
Material.

The calculations consist of two steps: (i) self-consistent DFT
calculation with the Wien2k code \cite{wien} and (ii) solving the
time-dependent Liouville equation. Since the helicity dependence of
spin reversal is always superimposed on the demagnetization (see
details in the Supplementary Material), we subtract the average spin
moment $\bar{M}_z=(M_z^{\sigma^+}+ M_z^{\sigma^-})/2$ from the moment
for each helicity to get the net effect of the helicity-dependence,
$\Delta M_z^{\sigma^{\pm}}=M_z^{\sigma^{\pm}}-\bar{M}_z$.  Here
$M_z^{\sigma^{+/-}}$ is the spin moment under $\sigma^{+/-}$
excitation.

\begin{figure}
\includegraphics[angle=0,width=0.8\columnwidth]{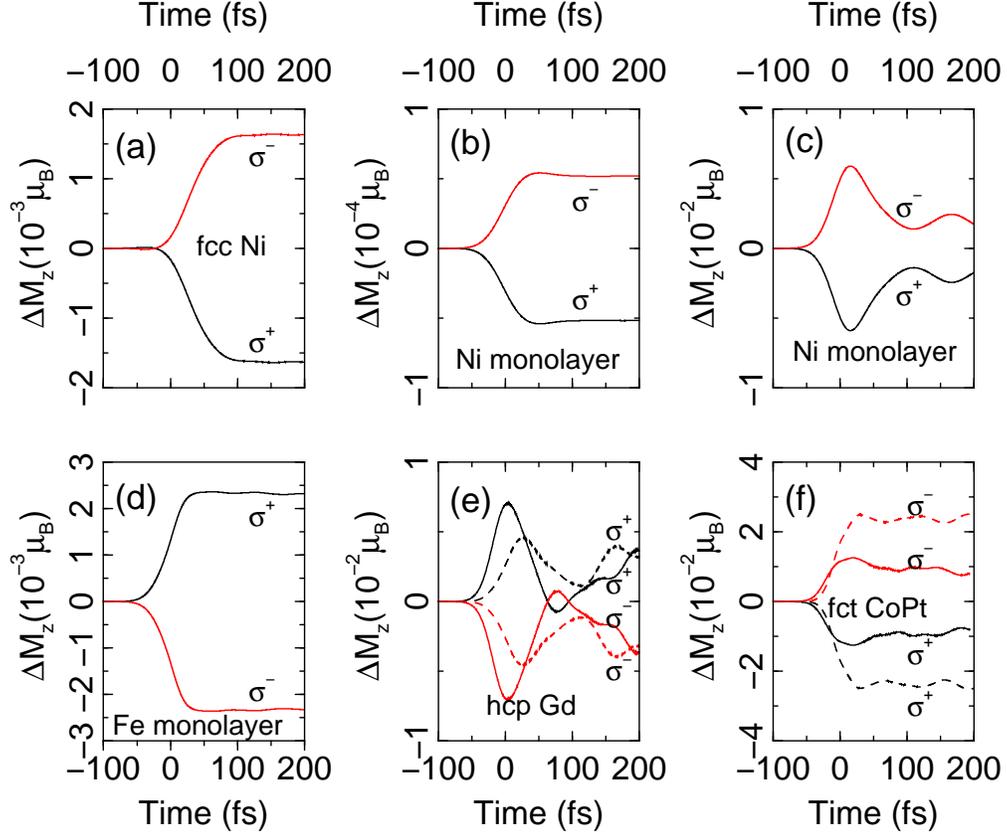}
\caption{ First-principles simulation of helicity-dependent spin
  moment change $\Delta M_z$ for (a) fcc Ni, (b) and (c) Ni
  free-standing monolayer, (d) Fe monolayer, (e) hcp Gd and (f) fct
  CoPt. $\Delta
  M_z^{\sigma^{\pm}}=M_z^{\sigma^{\pm}}-(M_z^{\sigma^{+}}+M_z^{\sigma^{-}})/2$.
  Here $M_z^{\sigma^{+/-}}$ is the spin moment under $\sigma^{+/-}$
  excitation.  Laser parameters are as follows.
(a) Duration $\tau=60$ fs, photon energy $\hbar \omega =2.0$eV and
  vector potential amplitude $A_0=0.0099 \rm V fs / \AA$.
(b)  $\tau=48$ fs, $\hbar \omega =1.6$eV and $A_0=0.0030 \rm V fs /
  \AA$.
(c) $\tau=48$ fs, $\hbar \omega =1.55$eV and $A_0=0.030 \rm V
  fs / \AA$.
(d) $\tau=48$ fs, $\hbar \omega =2.0$eV and $A_0=0.030
  \rm V fs / \AA$.
(e) (solid line)  $\tau=48$ fs, $\hbar \omega =1.6$eV
  and $A_0=0.030 \rm V fs / \AA$.
(dashed line) $\tau=48$ fs, $\hbar \omega =1.55$eV
  and $A_0=0.030 \rm V fs / \AA$.
(f) (solid line) $\tau=48$ fs, $\hbar \omega =1.6$eV and $A_0=0.030
  \rm V fs / \AA$. (dashed line) $\tau=48$ fs, $\hbar \omega =1.55$eV
  and $A_0=0.030 \rm V fs / \AA$.   }
\label{mono3}
\end{figure}

Figure \ref{mono3}(a) shows that $\sigma^+$ and $\sigma^-$ have
different effects on the moment in fcc Ni, and $\sigma^+$ reduces the
moment more, which can be understood from the above dipole selection
rule.  Such a helicity dependence is also observed in a Ni free-standing
monolayer (see Fig. \ref{mono3}(b)).  If we increase the field
amplitude by ten times, the moment change becomes oscillatory for
$\sigma^+$ and $\sigma^-$ (see Fig. \ref{mono3}(c)), and the net change
in moment increases 100 times, but the relative moment change for
$\sigma^+$ and $\sigma^-$ remains the same.

However, this is no longer the case for a Fe monolayer, where $\sigma^-$ decreases the moment more than $\sigma^+$
(see Fig. \ref{mono3}(d)).  This is because only those pockets in the
$k$ space that are optically accessible can contribute to the moment
change, and the global moment direction may not align with the local
moment direction.

hcp Gd is particularly interesting. The solid line in
Fig. \ref{mono3}(e) shows that $\sigma^-$ induces a larger change, but
around 75 fs, $\sigma^+$ has a larger change. If we reduce the photon
energy to 1.55 eV, such a crossover is not seen (see the dashed
lines in Fig. \ref{mono3}(e)). We did not find a similar case in other
materials investigated.

fct CoPt is very unique and has a strong magnetic anisotropy. Its
moment change (Fig. \ref{mono3}(f)) is larger than others under a
similar laser excitation, and increases twice if we use a 1.55-eV pulse
instead of 1.60-eV. This reflects the importance of the laser photon
energy.  In summary, our first-principles result unambiguously
demonstrates that the helicity-dependence of the moment dynamics is
generic, but the degree of the helicity effect is very much
material-dependent.

\subsection{Impact of sublattice spins on spin reversal }

While our first-principles investigation lays the ground work for spin
reversal, it can not fine-tune its sublattice spins at each lattice
and give no direct information about the effect of sublattice spin
ordering on spin reversal.  Our model, with a realistic laser pulse,
provides complementary information.  We investigate three
representative spin systems -- ferromagnetic (FM), weak and strong
ferrimagnetic (FIM) -- to approximately simulate ferromagnets
\cite{lambert,mangin}, TbCo alloys
\cite{sabineprb14,hassdenteufel2015}, and to some extent GdFeCo alloys
\cite{stanciu}, respectively.  We emphasize that all the calculations
below use the same sample geometry ($21\times 21\times 2$) and
parameters, and we change only the sublattice spins, so their impact
on spin reversal can be investigated unambiguously.

\begin{figure}
\includegraphics[angle=0,width=0.8\columnwidth]{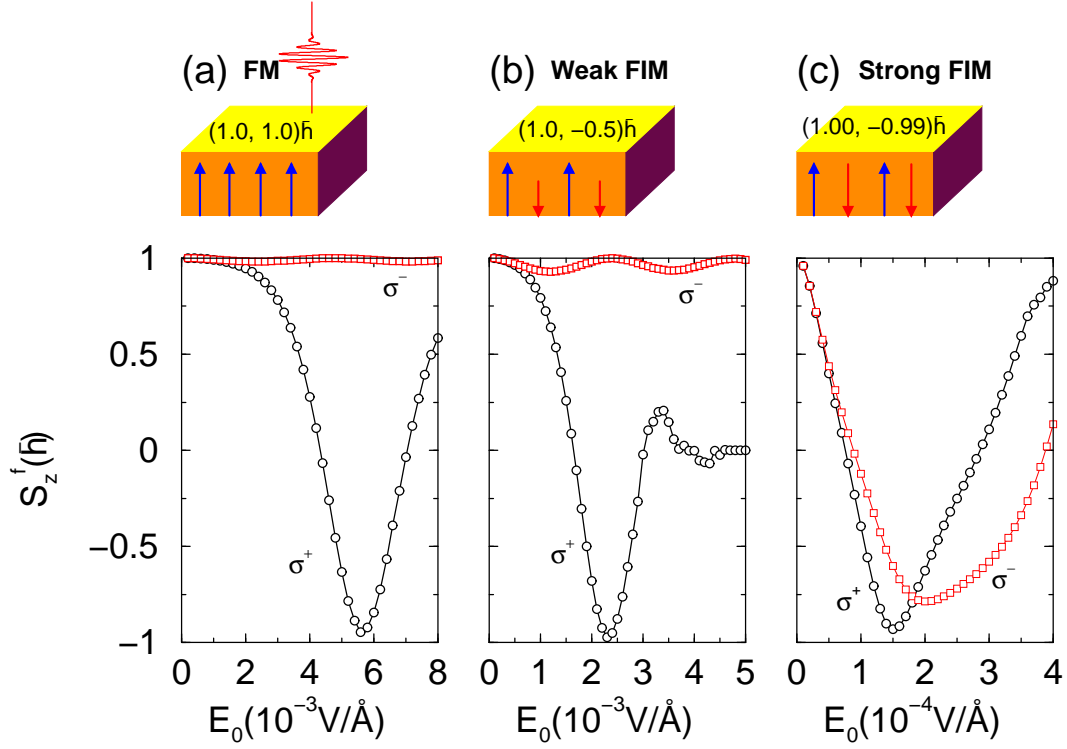}
\caption{ Dependence of the final spin angular momentum on the laser field
  amplitude $E_0$ under right- and left-circularly polarized light in a
  (a) ferromagnet, (b) weak ferrimagnet and (c) strong
  ferrimagnet. $\tau=240$ fs.  The empty circles denote the results
  with $\sigma^+$ and the empty boxes those with $\sigma^-$. The
  optimal amplitudes for $\sigma^+$/$\sigma^-$ reduce from (a)
  0.0056/0.0024 $\rm V/\AA$, (b) 0.0023/0.0012 $\rm V/\AA$ to (c)
  0.00015/0.0002 $\rm V/\AA$.  }
\label{helicity}
\end{figure}

Figure \ref{helicity} shows a comprehensive view how the spin reversal
depends on the laser field amplitude $E_0$ as the spin ordering
changes from a ferromagnetic to weak and then strong ferrimagnetic
phase. Figure \ref{helicity}(a) shows that the ferromagnetic layer,
with a single sublattice spin $S^a_z=S^b_z=1\hbar$, has a pronounced
helicity dependence.  With the initial spin up, only $\sigma^+$ is
effective, and $\sigma^-$ virtually has little effect, fully
consistent with the selection rule discussed above and the
experimental findings \cite{lambert}.

The situation is different when the system has two spin
sublattices. As seen in Fig. \ref{fig0}, when both spin orientations
are present, $\sigma^+$ and $\sigma^-$ excite different sets of spins.
We retain the spin on sublattice $a$ but flip and reduce the spin on
sublattice $b$ by half to $S^b_z=-0.5\hbar$, which we call a weak
ferrimagnet, thus mimicking TbCo alloys.  We note in passing that
these spin angular momenta are chosen as examples and have no effect
on our conclusion qualitatively, as far as they satisfy the minimum
momentum requirement \cite{epl16}.  Figure \ref{helicity}(b) shows
that $\sigma^+$ is capable of reversing the spin from $1.0\hbar$ to
$-0.98\hbar$ (see the empty circles), with nearly 100\% switchability,
even with a smaller optimal field amplitude of $2.3\times 10^{-3}\rm
V/\AA$ than $5.6\times 10^{-3}\rm V/\AA$ in the FM case. In contrast
to FM, down spins on sublattice $b$ of FIM, which are not supposed to
directly flip under $\sigma^+$ according to Fig. \ref{fig0}, are also
switched over but {\it indirectly through the exchange interaction
  $J_{ex}$}.  This means that unlike FM, FIM has two channels to
switch spins, either directly through {\it correct} light-helicity or
indirectly through the exchange interaction.  $\sigma^-$ also affects
the spins, and there is a clear modulation in the spin with the field
amplitude (see the empty boxes in Fig. \ref{helicity}(b)), but
$\sigma^-$ is still much less effective.  This reveals a crucial
insight that if the sublattice spin magnitudes differ a lot, only one
helicity can reverse spins effectively, so the switching remains
highly helicity-dependent. We believe that this is what happens in
ferrimagnetic TbCo.  Its sublattice effective spin on Tb is much
larger than that on Co. Here the effective spin must be used since
TbCo alloys have different concentrations \cite{epl16,mplb16}.

So far, all the switchings are helicity-dependent. To understand HIDS,
we need to understand the magnetic structure difference between GdFeCo
and TbCo alloys. Fe has a larger spin moment than Co, so the effective
spin difference between Gd and Fe sites in GdFeCo is much smaller than
that between Tb and Co in TbCo. In fact, GdFeCo alloys have nearly
compensated spins on two sublattices.  Hassdenteufel \et
\cite{hassdenteufel2015} even proposed the low remanence as the
criterion for AOS.  Our strong ferrimagnet model simulates such a
scenario where the spin on sublattice $b$ is only 1\% smaller than the
spin on sublattice $a$, i.e. $S^a_z=1\hbar$ and $S^b_z=-0.99\hbar$,
(see Fig. \ref{helicity}(c)).  The system is very close to an
antiferromagnet. Figure \ref{helicity}(c) shows that both $\sigma^+$
and $\sigma^-$ are effective to switch spin, thus realizing a
helicity-independent switching.  $\sigma^-$ induces a final average
spin of $-0.79\hbar$ at the optimal field amplitude.



\begin{figure}
\includegraphics[angle=0,width=0.8\columnwidth]{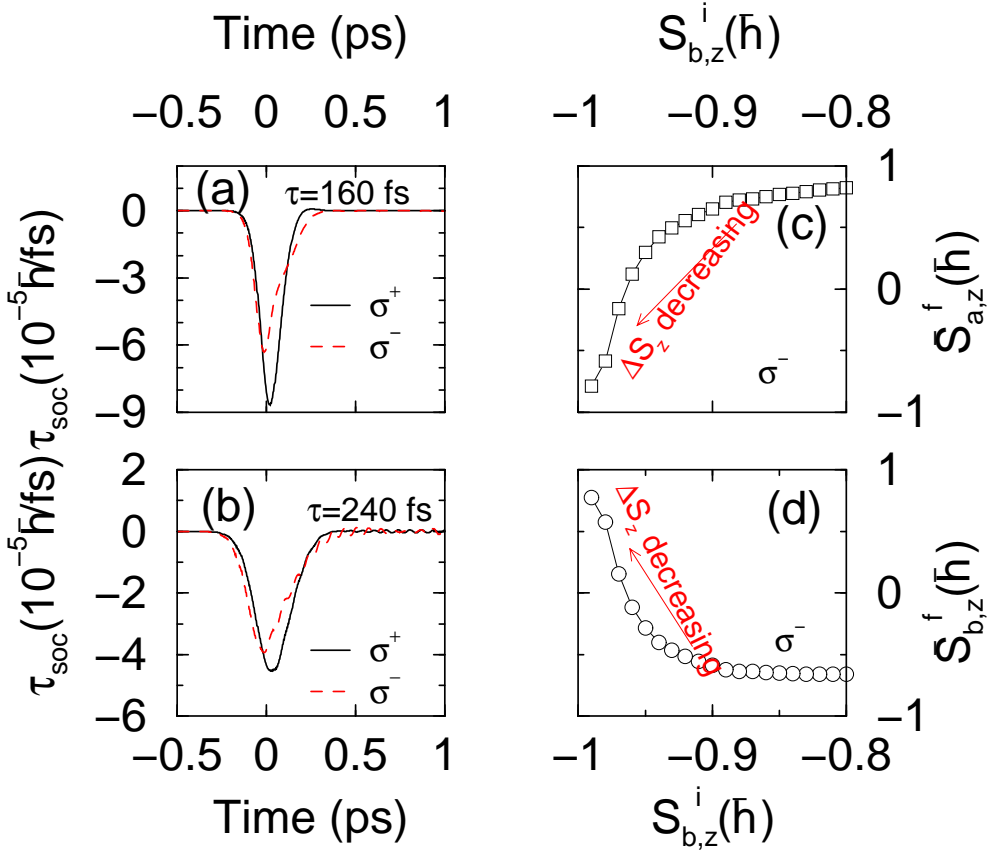}
\caption{(a) Laser-induced spin-orbit torque $\tau_{soc}$ as a
  function of time for $\sigma^+$ (solid line) and $\sigma^-$ (dashed
  line) pulses; $\tau=160$ fs.  (b) Same as (a) but $\tau= 240$ fs.
  (c) Final average spin at sublattice $a$ as a function of the
  initial spin on sublattice lattice $b$. $\Delta S_z$ is the
  sublattice spin magnitude difference.  (d) Final average spin at
  sublattice $b$ as a function of the initial spin on sublattice
  lattice $b$.  }
%
\label{torque}
\end{figure}

The importance of sublattice spins has long been recognized
\cite{mentink}, but the interplay between the sublattice spin and
light helicity is not.  In Fig. \ref{torque} we explain why $\sigma^+$
appears more powerful to reverse spins than $\sigma^-$.  Figure
\ref{torque}(a) shows the time evolution of optical spin-orbit torques
(OSOT) \cite{epl16} for $\sigma^+$ (solid line) and $\sigma^-$ (dashed
line) pulses.  The electric field in Eq. (\ref{efield}) first excites
the orbital angular momentum \cite{mplb16} and then OSOT.  The
definition of OSOT is $\tau_{\rm soc}=\lambda {\bf L} \times {\bf S}$.
It is clear that OSOT critically depends on the magnitude of the spin
(compare the solid and dashed lines), and the spin evolution contains
both precession and flipping.  Since the down spin has a smaller
magnitude, its torque is smaller, so the switching under $\sigma^-$
excitation is not as perfect as that under $\sigma^+$.  Increasing the
pulse duration from $\tau=160$ to 240 fs (Fig. \ref{torque}(b))
reduces the torque difference between $\sigma^+$ and $\sigma^-$.
These torques are the time-dependent analogue of the effective
magnetic field introduced in the inverse Faraday effect (IFE)
\cite{stanciu}, but IFE has never been formulated in terms of light
helicity and spin and orbital angular momenta \cite{pop}, so it is
unable to draw the crucial connection to AOS.  Our finding establishes
an important paradigm that sublattice spins directly impact how the
helicity switches spins, being helicity-dependent or
helicity-independent.  The key is that the spin-orbit torque
$\tau_{soc}$ intrinsically depends on the magnitude of the spin, thus
the helicity-dependence of AOS becomes spin-dependent.  Figure
\ref{torque}(c) shows that as $S^i_{b,z}$ decreases from $-0.8\hbar$
to $-0.9\hbar$, the final spin on sublattice $a$, $\bar{S}_{a,z}^f$,
has a very small decrease, but once $S^i_{b,z}$ is below $-0.95\hbar$
or $\Delta S_z$ is below $0.05\hbar$, $\bar{S}_{a,z}^f$ decreases
superlinearly and reaches $-0.79\hbar$.  The results for sublattice
$b$ are plotted in Fig.  \ref{torque}(d). We see similarly that as
$\Delta S_z$ decreases, $\bar{S}_{b,z}^f$ increases superlinearly.

\subsection{Snapshot of spin reversal}

\begin{figure}

\includegraphics[angle=0,width=0.7\columnwidth]{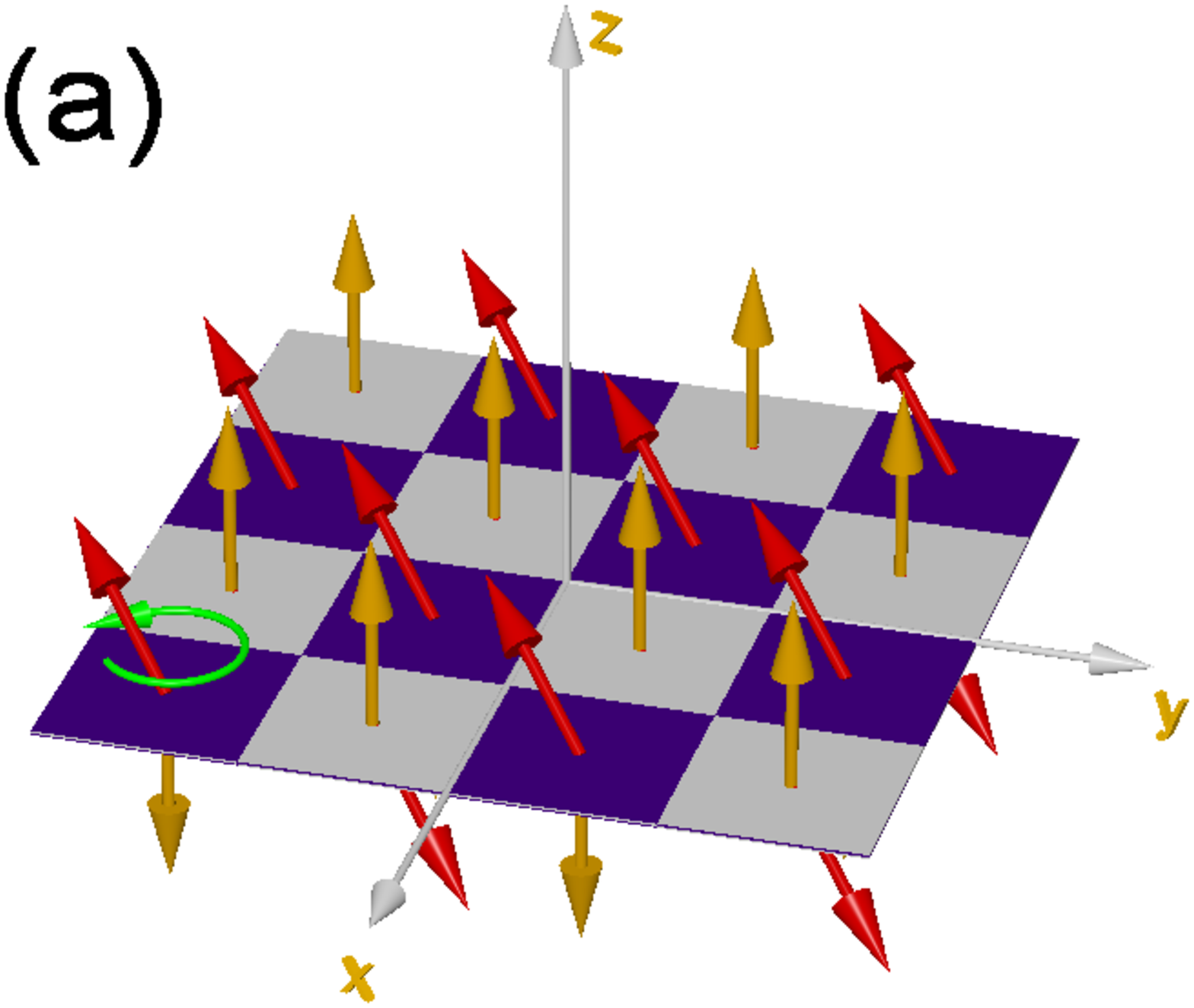}
\includegraphics[angle=0,width=0.7\columnwidth]{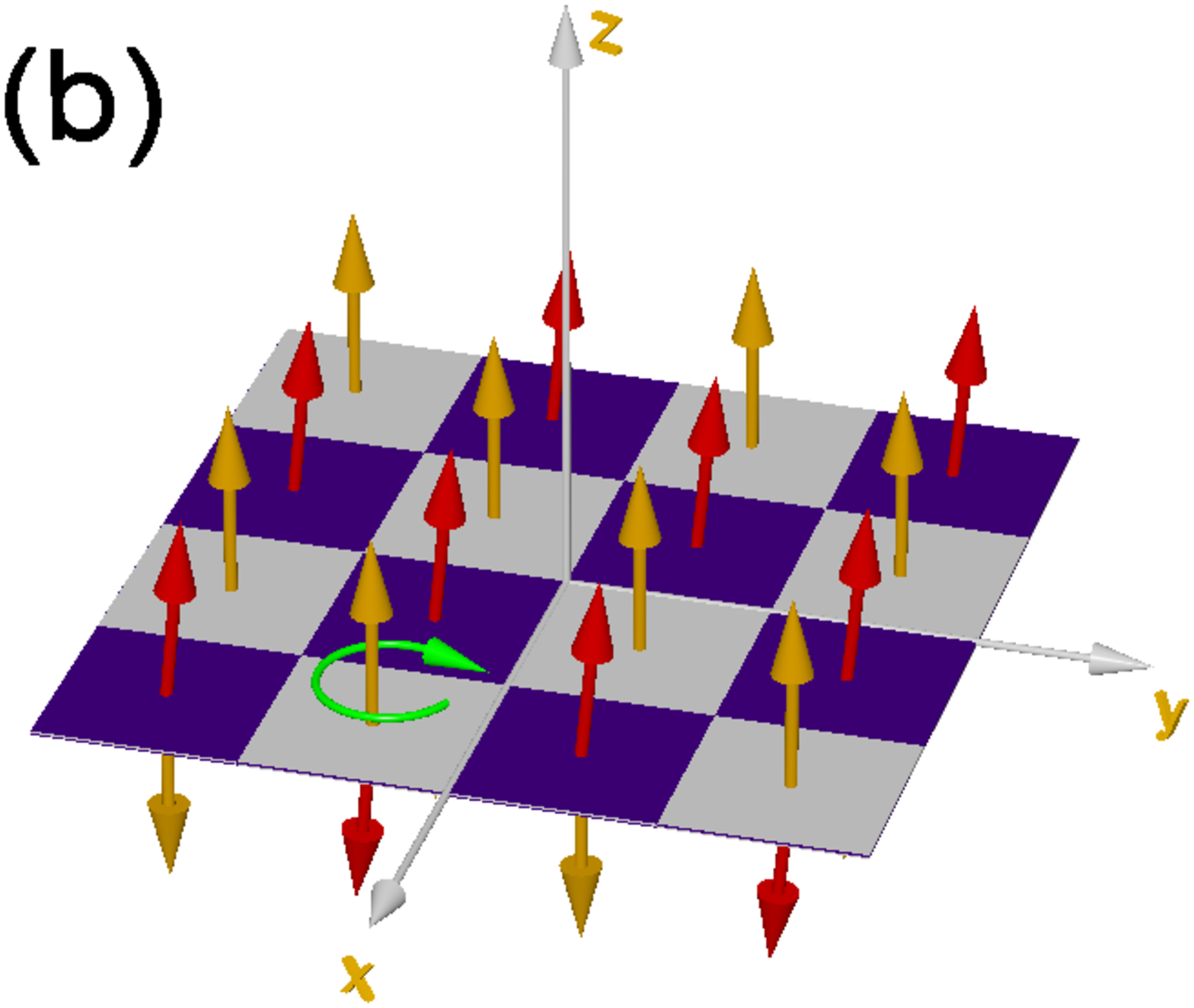}
\caption{ (a) Snapshot of spins at the center of a strong
  ferrimagnet  before (gold arrows) and 2 ps
  after (red arrows) $\sigma^-$ excitation.  A partial reversal is
  observed.  The torus arrow shows which spins the laser initially
  switches.  (b) Snapshot of spins at the center of the slab before
  (gold arrows) and 2 ps after (red arrows) $\sigma^+$ excitation. A
  nearly complete reversal is found. The torus arrow shows which spins
  the laser initially switches.  }
\label{allspin}
\end{figure}

So far we have shown the spin dynamics of one representative spin. Now
we show a group of spins at the center part of the first layer.  The
gold arrows in Fig.  \ref{allspin}(a) are the initial spins on two
sublattices. They take values of $+1\hbar$ and $-0.99\hbar$ and form a
ferrimagnetic network extending along all three directions.
The red
arrows in Fig.  \ref{allspin}(a) capture a snapshot of the spins at 2
ps after $\sigma^-$ excitation.  Spins in the second layer are similar
(not shown).
  We see
that all the spins, regardless of their original orientations, are
reversed, or more precisely, cant toward opposite directions.  The green
torus arrow highlights that a $\sigma^-$ pulse selectively switches
those down spins up first and then those initial up spins through the
exchange interaction.  Figure \ref{allspin}(b) shows that the
switching with $\sigma^+$ is nearly perfect, with all the spins
pointing in the opposite directions of the initial spins.  The green
torus arrow shows another example that a $\sigma^+$ pulse selectively
switches the up spins first.  If the white squares beneath in
Fig. \ref{allspin}(a) represent spin up and the blue ones down,
$\sigma^-$ and $\sigma^+$ are going to reverse those domains
selectively.

The entire process is pretty much similar to the super-resolved
fluorescence microscopy which was recognized by the Nobel Prize in
Chemistry in 2014, where fluorescent proteins act as an agent to beat
the diffraction limit. In our case, the agent is the magnetic
domain. They allow sub-wavelength imaging, which may explain ultrafine
magnetic domains created in experiments \cite{mangin}.  Our above
finding also explains why experimentally El Hadri \et \cite{elhadri}
found that $\sigma^+$ ($\sigma^-$) switches the magnetization to down
(up). In their Co-dominated TbCo alloy films, the spins at Co sites
point up initially, while in Tb-dominated films, the spins are
down. These two different experiments beautifully demonstrate how
accurate our prediction is.


\subsection{Phase diagram }

\begin{figure}

\includegraphics[angle=0,width=0.7\columnwidth]{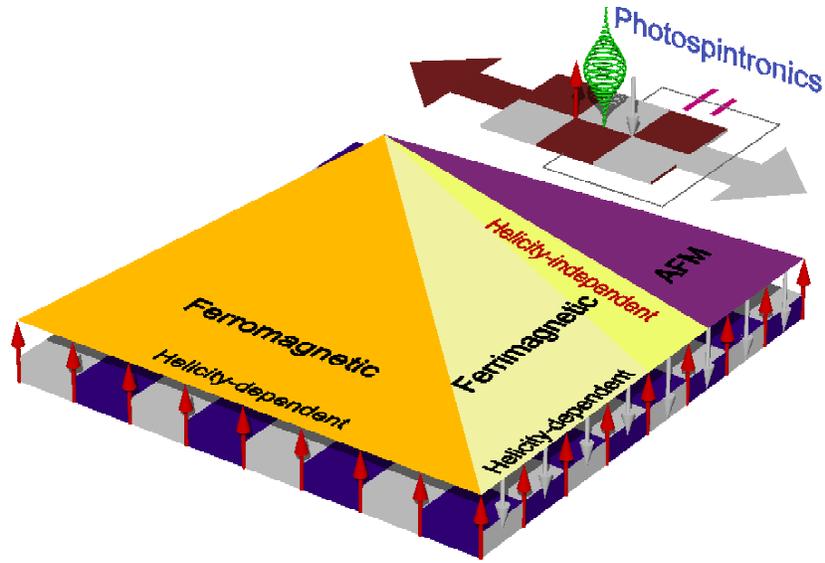}
\includegraphics[angle=0,width=0.7\columnwidth]{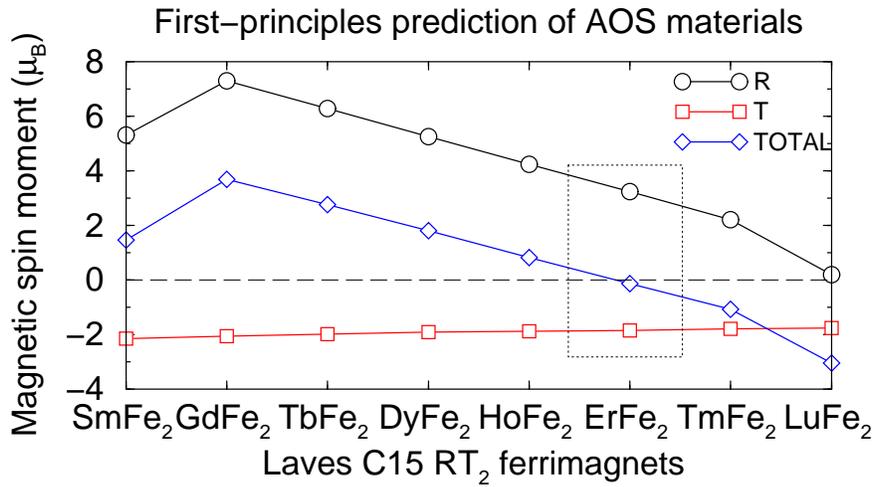}
\caption{ (Top) Phase diagram of AOS. Switchings in FM (orange
  triangle) and weak FIM (light yellow triangle) are always
  helicity-dependent. Helicity-independent switching (yellow triangle)
  occurs in a narrow region when the sublattice spins approach the
  antiferromagnetic limit.  (Top right) The envisioned photospintronic
  device is based on a strong ferrimagnet which allows the laser to
  store and switch spins rapidly. (Bottom) Magnetic spin moment for
  eight Laves phase C15 rare-earth-transition metal ferrimagnets. The
  spin moment at Fe site is almost constant, but that at R site peaks
  at $\rm GdFe_2$ and decreases along the series. Around ErFe$_2$,
  there is an optimal strong ferrimagnetic configuration (dashed box)
  where an ideal spin reversal may appear. The dashed line is at 0
  $\mu_B$.  }
%
\label{phase}
\end{figure}

Based on the above results, we construct a phase diagram for AOS in
Fig. \ref{phase}.  The essence of this phase diagram is that all the
AOS materials should be classified into three types: ferromagnets,
weak ferrimagnets and strong ferrimagnets.  On the left, we show that
AOS in ferromagnets such as CoPt \cite{lambert} is always
helicity-dependent (see the orange triangle). AOS in ferrimagnets (the
light yellow triangle) such as TbCo \cite{sabineprb}, where the
sublattice spins differ a lot, is also helicity-dependent. A sudden
change occurs when the sublattice spins differ very little from each
other in strong ferrimagnets such as GdFeCo, just before they become
antiferromagnetic. Regardless of laser helicity, the switching is
possible for both helicities. The yellow triangle denotes this region.
This phase diagram does not only unify paradoxically different
switching theories onto two simple concepts -- the optical selection
rule and the sublattice spin difference -- but it also suggests a
practical protocol for experimentalists.

Since an expensive sub-100 fs laser would limit the wide application of
AOS, we also examine whether a longer pulse can switch spins as well.
Table \ref{table} shows that as $\tau$ increases from 160 to 480 fs
\cite{chen} the optimal amplitude is significantly reduced as
expected. What is even better is that the switching becomes more
robust, with a more negative final spin and a much smaller
peak-to-peak amplitude $\delta$. This points out an effective path to
integrate the ultrafast magnetic storage into rapid optical switching
for communication by using stronger ferrimagnets and reasonably longer
laser pulses.  On Oct. 4, 2016, Peregrine Semiconductor Corp announced
60 GHz switches and with 8 ns switching time \cite{pere}.  Here we see
that in the strong ferrimagnet pumped with a 480-fs pulse, the spin
reversal time $T_r$ is 609 fs, or 1.6 GHz, thus easily beating the
above record.  In the top right of Fig. \ref{phase}, we envision an
integrated photospintronic device, where an ultrafast circularly
polarized laser pulse stores magnetic bits into a ferrimagnet, and
the medium controls the signal switching. The signal can be picked up
through electric circuits.

\section{Future applications}

The state-of-the-art energy consumption for telecommunication is 10 fJ
\cite{miller}.  Figures \ref{helicity}(a) through \ref{helicity}(c)
show that for the same laser parameter, FM needs a much stronger field
on the order of $10^{-3}\rm V/\AA\propto 10\rm MW/cm^2$, but the power
drops to 0.1 $\rm MW/cm^2$ in strong FIM (see Fig.
\ref{helicity}(c)). We use the experimental parameters from Chen \ete \cite{chen} and find that the energy consumption is 0.3 fJ. This
already meets the requirement of telecommunication switching
\cite{miller}. Therefore, tailoring FIM toward an even stronger FIM is
likely to accelerate the deployment of AOS-based switching technology.

We can move one step further to suggest some new candidates for AOS.
The bottom of Fig. \ref{phase} shows the computed spin moments at each
rare-earth side and Fe for eight Laves phase C15 phase alloys (RT$_2$) from
SmFe$_2$ through LuFe$_2$ (the details of the calculation are
presented in the Supplementary Material). We see that early in the
lanthanide series the spin moment on R is much larger than Fe and
peaks at GdFe$_2$. This explains why in amorphous GdFeCo the
concentration of Gd must be low. However, as in the latter part of the
series, the spin moment decreases, so crystalline RT$_2$ becomes a
strong ferrimagnet. A dashed line box around ErFe$_2$ highlights such
a case. Experimentally, growing these materials has gained renewed
interest \cite{lee,williams}. It is our belief that our finding will
further motivate and ignite intense research on photospintronic
applications.

\section{Conclusions}

We have carried out a joint time-dependent first-principles and model
calculations to pin down an alternative origin in thirteen different
magnetic systems. Our results show that all-optical spin switchings
can be unified under two crucial concepts: the optical selection rule
and the sublattice spin difference. The selection rule dictates that
left-(right-)circularly polarized light only switches the spin from
down (up) to up (down).  This one-to-one correspondence between spin
orientation and light helicity is generic, as confirmed by our
first-principles results.  We construct a phase diagram to categorize
all the magnetic materials into three categories.  In ferromagnets,
only one spin orientation is present, so that they show a strong
helicity-dependent switching. For ferrimagnets, we need the second
concept -- sublattice spin difference. For weak ferrimagnets, with
very different sublattice spins, the switching is also
helicity-dependent. For strong ferrimagnets, with similar sublattice
spins, the switching becomes helicity-independent, and both $\sigma^+$
and $\sigma^-$ can reverse spins.  This conclusion is independent of
the system size and exchange interaction, and is fundamental to AOS.
This represents a paradigm shift for AOS and may have a far-reaching
impact on the future of fast magnetic storage technology. We compute
the energy consumption in those optimal ferrimagnets and find that it
already meets the requirements of the current technology. We have
further studied a group of Laves phase C15 rare earth alloys and find that
their spin moments are ideal for real devices. We expect that our
results will motivate further investigations into the laser-induced spin
reversal.

\acknowledgments We would like to thank Drs. J. Y. Chen, J. P. Wang
and M. Li of University of Minnesota for helpful discussions.  This
work was solely supported by the U.S. Department of Energy under
Contract No. DE-FG02-06ER46304. Part of the work was done on Indiana
State University's quantum cluster and high-performance computers.
The research used resources of the National Energy Research Scientific
Computing Center, which is supported by the Office of Science of the
U.S. Department of Energy under Contract No. DE-AC02-05CH11231.








\begin{table}
\caption{ Dependence of spin switching on the laser pulse duration in a
  ferrimagnetic ordered slab under $\sigma^+$ pulse excitation.  The
  system size is $21\times 21\times 2$. Spins on sublattices $a$ and
  $b$ are $1\hbar$ and $-0.99\hbar$, respectively.  The exchange
  interaction is 0.1 eV/$\hbar^2$.  $E_{opt}$ denotes the optimal
  laser field amplitude, and $\bar{S}^f$ is the final time-averaged
  spin at sublattice $a$. $\delta$ is the peak-to-peak amplitude.  The
  spin reversal time $T_{r}$ is defined as when the spin
  reaches its first minimum.  }
\vspace{1cm}
\begin{tabular}{cccccc}
\hline\hline
$\tau$ (fs) &  $E_{opt}$ (V/$\rm \AA$) & $\bar{S}^f(\hbar)$ & $\delta
  (\hbar)$ &  $T_{r}$(fs) \\
\hline
160   & $2.9\times 10^{-4}$ & -0.82 & 0.36 &218.09\\
200   & $2.0\times 10^{-4}$ & -0.90 & 0.20 &305.33\\
240   & $1.5\times 10^{-4}$ & -0.93& 0.14 &392.57\\
360   & $0.9\times 10^{-4}$ & -0.94 & 0.10 &544.42\\
480   & $0.7\times 10^{-4}$ & -0.95 & 0.07 &609.04 \\
\hline\hline
\label{table}
\end{tabular}
\end{table}

\end{document}